# Defect induced magnetism and super spin glass state in reactive ion beam deposited nano structured AlN thin films

Shilpam Sharma[1], E. P. Amaladass[1], Awadhesh Mani[1*]

[1]Condensed Matter Physics Division, Materials Science Group, Indira Gandhi Centre for Atomic Research, HBNI, Kalpakkam 603102, India

Defect induced magnetism is reported in undoped aluminium nitride thin films deposited using reactive ion beam sputtering of aluminium in nitrogen plasma. The films have been deposited on silicon substrate at different temperatures. Existence of the defects in the films have been verified using room temperature photo-luminescence measurements. Owing to nano crystalline nature of the films, super-magnetic ground states have been observed. A cross-over from super paramagnetic to super spin glass state has been observed as the grain density increases. Detailed magnetisation measurements along with AC susceptibility measurements have been used to determine the ground states in these AlN thin films.

**Keywords:** dilute magnetic semiconductors, ion beam sputtering, magnetisation, super spin glass

*Corresponding Author: Materials Science Group, Indira Gandhi Centre for Atomic Research,

Kalpakkam 603102, India.

Ph: 044-27480081

Email: mani@igcar.gov.in

## 1. Introduction

Aluminium nitride (AlN), an insulator having a wide direct band-gap ~ 6.1 eV, is a promising candidate in the field of dilute magnetic semiconductors (DMS) for spintronics applications [1-8]. Previously, doping of a few percent of magnetic transition metal (TM) atoms such as Cr, Mn, Fe, V and Ni has been utilized to produce AlN based DMSs [1, 9-11]. However, in the case of magnetic dopants, there is always an ambiguity about the origin of magnetic order in DMS. In addition to the intrinsic magnetic order, the dopant cluster or secondary phases may also give rise to magnetism in an otherwise non-magnetic semiconductor host. Owing to their small concentrations, these magnetic precipitates can easily escape their detection. To avoid such uncertainties, non-magnetic dopants or native defect induced magnetism offer a better choice to achieve magnetic semiconductors. It has been demonstrated in several experimental and theoretical works that doping of non-magnetic Mg, Si and Cu favour a ferromagnetic ground state in AlN bulk and nano-structures [7, 12-15]. Similarly, it is well established that the nitrogen and/or aluminium vacancies prefer spin polarisation and can give rise to magnetic order beyond certain concentration [16-18]. The native defects have been shown to even enhance the magnetisation in TM doped AlN matrix [5, 8, 19]. Unlike in AlN, the native vacancy defects does not prefer spin polarisation in the two dimensional $MoS_2$ [20] which is another promising candidate for spintronics applications. Among the two prominent native defects: anionic and cationic vacancies, the cationic or Al vacancy ($V_{Al}$) is shown to produce large spin polarisation [16]. However, generation of Al-defects is more difficult than N- vacancies ($V_N$) in nano-sheets, thin film or bulk form [21] and thus in most of the reports, magnetisation in defected AlN has been attributed to the nitrogen vacancies [17]. In this context, it is vital to note that the AlN thin films deposited via ion beam sputtering of aluminium in reactive assistance of nitrogen plasma are susceptible to the generation of vacancy type native defects [22] and thus can provide a testbed for the studies on defect induced magnetism in undoped AlN thin films. Moreover, precise control over the properties of the reactive assistive nitrogen plasma can allow one to easily manipulate the defect density and grain sizes to play around with the final magnetic state of the undoped AlN based DMS system.

The DMS can be considered as the system of localized magnetic moments arbitrarily dispersed in a non-magnetic matrix of semiconductor. These localized spins get coupled to each other by either long range ferromagnetic/antiferromagnetic interactions, or can get frustrated and may land up in spin glass phase [23-25]. Upon decreasing the size of the grains below the critical radius of single magnetic domain, a large magnetic moment or super spin arises in the direction of easy axis. A variety of phases may then appear depending upon the concentration of these super spin entities. The secluded, non-interacting magnetic nano-particle can act as a super spin with super-paramagnetic (SPM) phase governing the dynamics of the system [26]. In SPM phase, the fluctuating super-spins at higher temperature gets blocked below a temperature $T_b$ where the relaxation time exceeds measurement time. SPM below $T_b$ has a ferromagnetic state and the temperature dependence of relaxation time obeys Arrhenius law. Similarly, a random distribution of strongly interacting and frustrated super spin ensemble freezes in to a super spin glass state (SSG) below a freezing temperature $T_g$ [26]. The relaxation time in the case of SSG exhibits a critical slowing down behaviour.

Here we report defect induced magnetisation and a super spin glass ground state in the undoped AlN thin films having grain sizes below 6 nm. The AlN thin films have been deposited at different substrate temperatures ($T_{dep}$ = 27-400 °C) using reactive ion beam sputter deposition of aluminium in nitrogen plasma. The presence of native defects in these films have been probed using room temperature photoluminescence (PL) spectra. A change from SPM state to SSG phase as the deposition temperature increases from room temperature to 400 °C has been observed for the first time in AlN thin film based DMS. Detailed measurements of DC magnetisation, AC susceptibility, ageing and relaxation have been performed on sample deposited at 400 °C and are presented.

## 2. Experimental Details

The AlN thin films have been deposited on Si(100) substrates by ion beam sputter deposition in reactive assistive mode by varying the substrate temperature from room temperature to 400 °C. Prior to deposition, the silicon substrates were cleaned using RCA1 process. A commercial 4" aluminium target of 99.999% purity was sputtered with argon ions extracted from a 6 cm RF ion source operated at 500 V, 80 mA fed with UHP argon of 99.9995% purity. Simultaneously a broad beam end-Hall type ion source was used to stream the reactive nitrogen ion flux directly to the substrate. The 200 mA reactive nitrogen ion flux was extracted with 90 eV energy from the assistive ion source. The reactive deposition was performed for 45 minutes. The base pressure in the deposition chamber was better than $2 \times 10^{-5}$ Pa and the working pressure was maintained at $3 \times 10^{-2}$ Pa using gas mass flow controllers. Thickness measurements were performed using Dektak profilometer and the thickness of all the films was found to be ~120 nm. X-ray diffraction measurements were performed to confirm phase purity of the samples and high resolution transmission electron microscopy (HRTEM) was used to find out the grain size and microstructure of the films. High resolution TEM measurements were performed using LIBRA 200FE HRTEM operated at 200 keV, equipped with an in-column energy filter and Schottky field emission gun source. Detailed sample synthesis and characterisation has been described elsewhere [27]. Presence of cationic and anionic vacancy type point defects was confirmed by photoluminescence (PL) spectroscopy with 325 nm He-Cd laser excitation using a Raman microscope (Renishaw plc, U.K., model inVia). Magnetisation and AC susceptibility measurements were performed on thin films deposited on rectangular substrates (2 mm x 5 mm x 500 micron thick) using Quantum design make MPMS-3 SQUID VSM under magnetic field up to 7 T and temperature varying from 2 K to 300 K.

## 3. Results and Discussion

The images of the HRTEM and small area electron diffraction (SAED) pattern for samples deposited at 100 ºC and 400 ºC are presented in Figure 1a and 1b respectively. It can be observed that the grain size and density increases with deposition temperature. The 100 ºC sample have grains of roughly 2-3 nm, embedded in an otherwise amorphous matrix. The grains are scarcely located in the sample as indicated by black circular regions in figure 1a. Inset in figure 1a shows the SAED pattern having broad rings which are signatures of amorphous materials along with few crystalline spots signifying a low density of small grains. In comparison the film deposited at 400 ºC has larger grains with 5-6 nm grain size. The particle size distribution of this film has been estimated by analysing dark field TEM micrographs using ImageJ software [28]. The average particle size has been estimated by fitting log-normal distribution to the size distribution histograms of roughly 100 different particles.

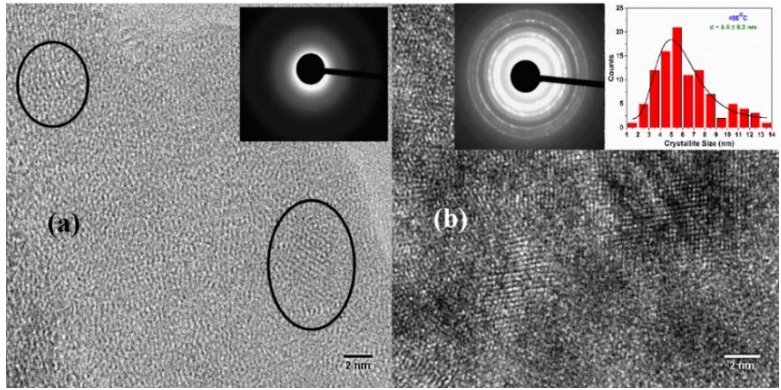

**Figure 1 a-b:** The HRTEM images of AlN thin films deposited at 100 and 400 °C shows the increase in grain size as well as grain density with increase in substrate temperature during deposition. The **inset of 1a** shows the SAED pattern of amorphous AlN film along with some crystalline spots. **Inset of 1b** shows the ring pattern from polycrystalline AlN film deposited at 400 °C. Inset also presents grain size distribution and its log-normal distribution which shows the average particle size of 5.6 nm.

The grain size distribution along with fit is shown in inset of figure 1b. The SAED pattern shown in figure 1b has well defined Debye-Scherrer rings indicating polycrystalline nature of the film. The *d*-spacings in SAED pattern were indexed to Wurtzite AlN and secondary or impurity phase were not observed. It has been recently reported that the surface properties, grain sizes and crystallographic orientation of AlN thin films deposited using reactive sputtering depends upon the substrate and metallic under layers [29].

The temperature variation of zero field cooled (ZFC) and field cooled (FC) magnetisation for samples deposited at room temperature, 300 and 400 °C is presented in figure 2. The samples were cooled to 2 K from 300 K in zero and 50 Oe field prior to recording ZFC and FC respectively. ZFC and FC measurements were performed while warming the sample from 2 K to 300 K under 50 Oe magnetic field applied parallel to film surface. The observed peak in ZFC magnetisation of all the samples corresponds to average blocking or freezing temperature of spin carrying AlN nano-particles. It has been observed that the blocking temperature increases with increase in the deposition temperature (cf. inset of fig.2) while, for a given sample, the blocking temperature decreases with increase in magnetic field (not shown). The FC magnetisation curve continues to increase with decrease in temperature but a small dip can be noticed in FC curve beyond the blocking temperature. The FC and ZFC curves bifurcate at a much higher temperature than the blocking temperature.

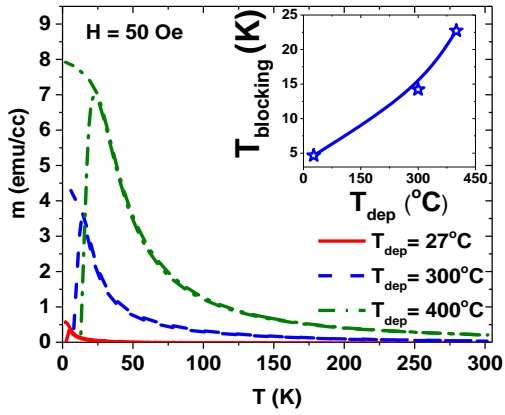

**Figure 2:** Zero field cooled and field cooled magnetisation as a function of temperature for AlN thin films deposited at different temperatures. The ZFC curve shows a peak below which magnetisation decreases monotonically. The FC curve bifurcates from ZFC at a temperature higher than the peak temperature indicative of the super-magnetic ground state in the samples. The peak temperature is seen to increase with increase in deposition temperature. **Inset:** The blocking temperature increases with the increase in the deposition temperature of AlN thin films.

Isothermal magnetisation M(H) loops recorded at 300 K and 2 K are presented in figure 3. Hysteresis appear in M(H) loops below the peak temperature. The coercive field for all the samples is 560-680 Oe but the saturation magnetisation increases with the deposition temperature. The presence of peak in ZFC, the bifurcation in ZFC and FC before blocking temperature ($T_B$) and hysteresis in M(H) below peak temperature are some of the indicators of magnetic nano-particle systems [26, 30].

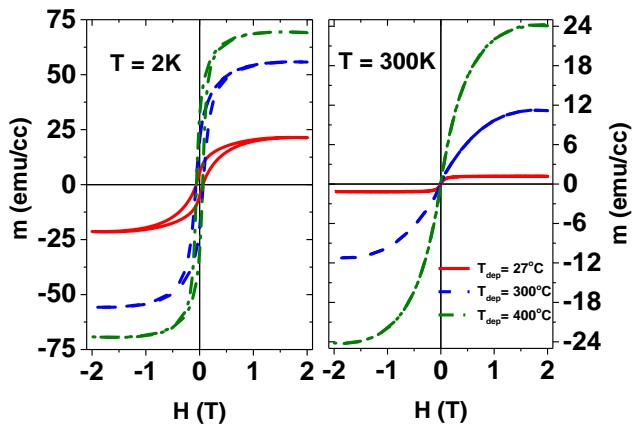

**Figure 3**: Isothermal magnetisation M(H) loops recorded at 300 K and 2 K shows hysteretic behaviour only below the blocking temperature. The coercive field for all the samples is 560-680 Oe but the saturation magnetisation increases with the deposition temperature.

No magnetic signals were observed in the measurements performed on the bare sample holder and 400 °C heat treated Si(100) substrate. This establishes that the magnetic signal has its origin in the AlN samples. Signatures of magnetic impurities could not be detected in X-ray photoemission spectroscopy measurements performed on the AlN thin films up to a depth of 50 nm [31].

The room temperature PL spectrum of AlN thin films deposited at different substrate temperatures has been presented in figure 4. Peaks corresponding to nitrogen and aluminium vacancies can be clearly seen in the PL spectrum. The broad peak centred on 2.95 eV corresponds to nitrogen vacancies [17] and a much smaller peak around 3.15 eV relates to aluminium vacancies in AlN [16].

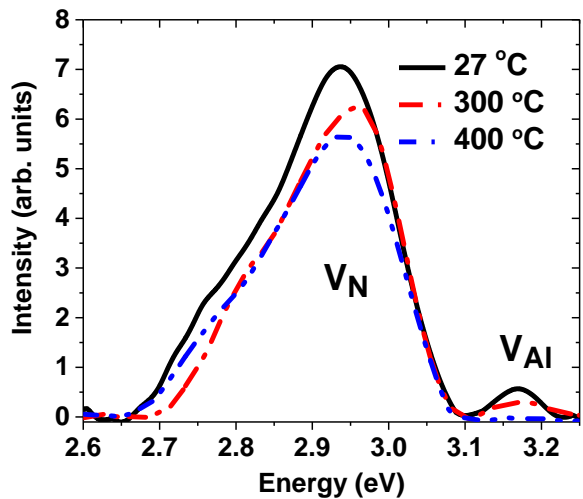

**Figure 4:** Room temperature photoluminescence spectrum of AlN thin films deposited at room temperature, 300 and 400 °C. The number of $V_{Al}$ is much smaller than the $V_N$ and both type of vacancies decrease with increase in deposition temperature.

It can be observed that the number of $V_{Al}$ is much smaller than the $V_N$ and both type of vacancy densities decrease with increase in deposition temperature. No signatures of $V_{Al}$ could be found in the sample deposited at 400 °C. This observation is in line with theoretical calculations which foretell that the formation energy of $V_{Al}$ is much larger than that of $V_N$ and thus the $V_N$ are found predominantly [12]. The existence of the native defects in these AlN thin films could be a possible reason for the observed magnetism in an otherwise diamagnetic AlN.

To further probe the state of the samples below their blocking temperature, AC susceptibility as a function of temperature has been measured at low excitation field ($H_{AC}$ = 2 Oe) with frequency varying between 1 Hz to 973 Hz. The data was collected under zero external field ($H_{DC}$ = 0 Oe) on ZFC samples. A representative plot of temperature variation of the real part of AC susceptibility normalized with its maximum value at each frequency are plotted in figure 5 for AlN thin film deposited at 400 °C. The peak height decreases and peaks shift towards higher temperatures with increase in the frequency of excitation field. This shift in peak temperature per decade change in frequency is quantified in the phenomenological parameter Ψ defined as $\frac{\Delta T_p/T_p}{\Delta log_{10}f}$. The parameter Ψ decreases with increase in the inter-particle interaction strength [32, 33]. In the case of non-interacting spins or SPM state, Ψ takes the value > 0.13, whereas for the spin glass state emerging from dipolar interactions between spin carrying particles $0.005 < \Psi < 0.05$ [32, 34].

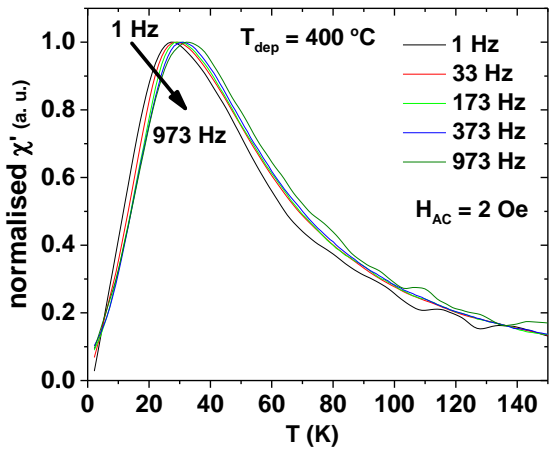

**Figure 5:** In-phase AC susceptibility as a function of temperature shows peak that shifts to higher temperature with increase in the frequency of AC excitation field.

The value of Ψ for the AlN thin film deposited at room temperature has been found to be 0.18 pointing to the SPM state below blocking temperature, but for the samples deposited at 300 and 400 °C, Ψ = 0.041 and 0.049 respectively thus indicating the presence of strong inter-particle interactions leading to super spin glass state below the freezing temperature.

To ascertain if the dynamics of the samples shows critical slowing down, the frequency dependence of peak temperature has been compared with the dynamic scaling theory which predicts that the relaxation time diverges at the glass temperature $T_g$. The temperature dependence of the relaxation time is given by the power law $\tau(f) = \tau^*(\frac{T_p}{T_g} - 1)^{-zv}$, where τ(f) is the frequency dependent spin relaxation time ($\frac{1}{2\pi f}$), $T_g$ is the glass transition temperature, τ* is the microscopic relaxation time related to the coherence of the coupled atomic spins in the nano particles and *zv* is the critical exponent[35]. Here, the freezing temperature $T_p$ has been taken as the temperature at which peak occurs in the real component of susceptibility, while $zv$, $\tau^*$ and $T_g$ are fitting parameters. The plots of relaxation time along with scaling power law fits for samples with $\Psi < 0.05$ is shown in figure 6. The value of critical exponent for samples deposited at 300 and 400 °C has been found to be 4.05 and 6.04 respectively. These values of critical exponent fall within the range for a glassy system ($zv \sim 4 - 12$) [34] thus indicating the existence of SSG state in these samples. The microscopic relaxation time $\sim 2 \times 10^{-7} - 8 \times 10^{-8}$ s obtained from the fit is also in line with the reported values for SSG states in different magnetic nano-particle systems [32, 36]. The $\tau(T_p)$ data of the sample deposited at room temperature could not be satisfactorily fitted to the scaling law thereby pointing to the absence of a critical slowing down in its spin dynamics. However, the $\tau(T_p)$ for that sample fitted very well to the Arrhenius law that describes the non-interacting single particle blocking across an energy barrier. The fitting to Arrhenius activation behaviour is shown in figure 6c and spin relaxation time of $\sim 6.1 \times 10^{-6}$ s with 33 K activation energy were found to be the best fit parameters for the room temperature deposited sample.

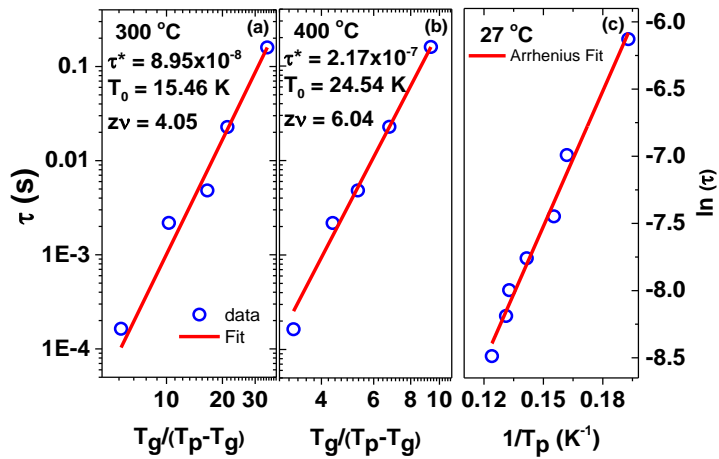

**Figure 6 (a-b):** The fitting of frequency dependent peak temperature to power law derived using dynamic scaling theory and value of exponent $zv$ shows that owing to the glassy state, the samples deposited at 300 and 400 °C have critical slowing down in their dynamics. The samples freezes in to a super spin glass state below peak temperature. **(c):** Arrhenius fit of frequency dependent peak temperature for sample deposited at room temperature points to the existence of non-interacting super paramagnetic ground state.

This analysis of in phase AC susceptibility data unambiguously points that the polycrystalline samples with inter particle interaction arising from larger grain density form a glassy state below their blocking temperatures, while the sample consisting of non-interacting, sparsely located grains attains a SPM state.

In order to confirm the spin glass freezing of spin carrying entities in AlN thin films, aging and memory experiments were performed under ZFC and FC protocols on sample deposited at 400 °C. It is known that the memory effect in FC protocol is observed in both SPM and SSG but in ZFC protocol it is observed in the case of SSG alone, thus providing an unambiguous signature of SSG state [37, 38]. In the FC protocol the sample was cooled from 300 K to 2 K under 50 Oe field with intermittent stops at 15, 10 and 5 K where the field was switched off for 120 minutes (7.2 kSec.), then returned to 50 Oe before resuming further cooling. Figure 7a shows the magnetisation curve while cooling with intermittent stops as $M^{IS}$. After field cooling the sample to 2 K, it was warmed up continuously up to 300 K under 50 Oe field. This field warming magnetisation is shown as $M^{FW}(T)$ in figure 7a. A reference FC magnetisation recorded while warming the sample under 50 Oe field is also shown in figure 7a. It can be seen that in the case of $M^{IS}$ the magnetisation falls during the wait time as the magnetic moments tends to equilibrate in zero field. The magnitude of the moment recovered after the field is applied back depends upon the response of magnetic moments to field. In the case of SSGs the spin dynamics show critical slowing down and thus the recovery of the moments is not efficient thereby producing a large drop in $M^{IS}$ curve. It can be observed that while warming the sample, the magnetisation shows a kink at every temperature of intermittent stops during cooling. This kink or memory of the stopping temperature occurs due to the recovery of lower energy magnetic configuration that the system remembers due to energy barrier redistribution during the cooling process [39, 40]. Beyond the freezing temperature, the magnetisation tracks back the reference FC magnetisation curve. A similar FC measurement with intermittent stops of zero applied field at 15, 10 and 5 K was also performed under 100 Oe field. The $M^{IS}$, $M^{FW}$ and reference FC magnetisation curves under 100 Oe field are plotted in figure 7b.

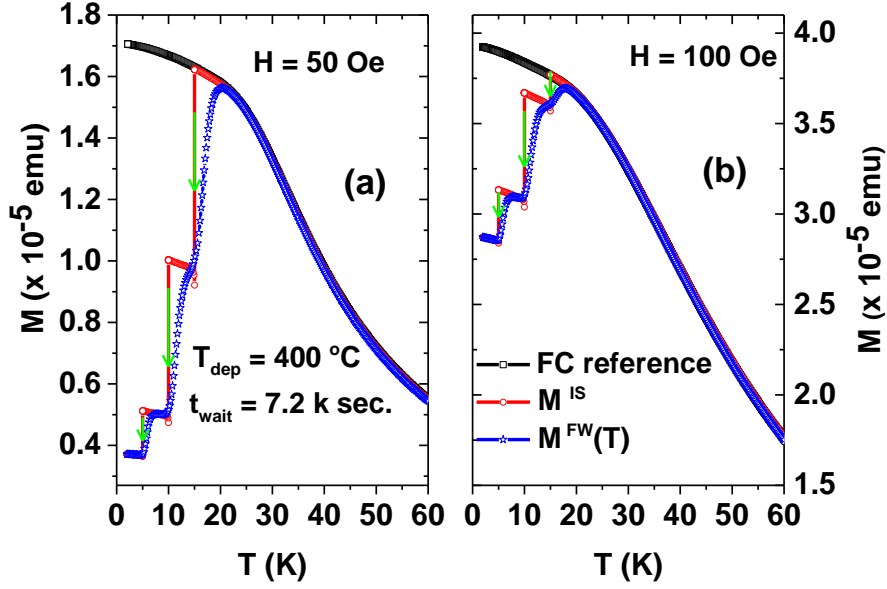

**Figure 7a:** Memory effect in FC DC magnetisation. The reference curve has been measured upon warming the FC sample in 50 Oe field. The $M^{IS}$ curve shows the magnetisation recorded during field cooling with intermittent stop and wait for 7.2 ks at 15, 10 and 5 K. After intermittent stops and cooling up to 2 K, the magnetisation measured during warming in 50 Oe field shows kinks at the temperatures where the cooling was intermittently stopped and sample was aged. **(b):** Similar experiment performed under 100 Oe field.

A memory effect in FC magnetisation is observed in both SPM and SSG states, but a similar effect in ZFC magnetisation is unique to the SSG systems. In the ZFC protocol, the sample in SSG state shows a dip at the temperature of stop and aging and the depth of the aging dip depends upon the wait time. For the time dependent memory experiment, the sample was first cooled to 15 K under zero magnetic field, aged for $5 \times 10^3$ and $3 \times 10^4$ s, then further cooled down to 2 K. At 2 K a small measuring field of 50 Oe was applied and magnetisation was recorded as a function of increasing temperature as the sample was warmed to 300 K. The reference ZFC magnetisation was recorded under same conditions but without any intermediate aging of the sample. The aging dip in the magnetisation manifest itself in the difference of DC susceptibilities between reference ZFC magnetisation and that with a single stop during cooling. The difference $\Delta\chi = \chi^{stop} - \chi^{ref}$ for different wait times is presented in figure 8a. It can be observed that the value of $\Delta\chi$ increases with increase in wait time. The ZFC magnetisation with a single stop for $10^4$ s, but at different temperatures 6, 9, 12 K was also recorded and the plots of $\Delta\chi$ as a function of temperature are presented in figure 8b. The dip in $\Delta\chi$ has been observed in the vicinity of the temperature at which the sample was aged during the wait time. This dip in the magnetisation of the sample in SSG state occurs due to relaxation of the system towards a more stable magnetic configurations during the stop and wait periods.

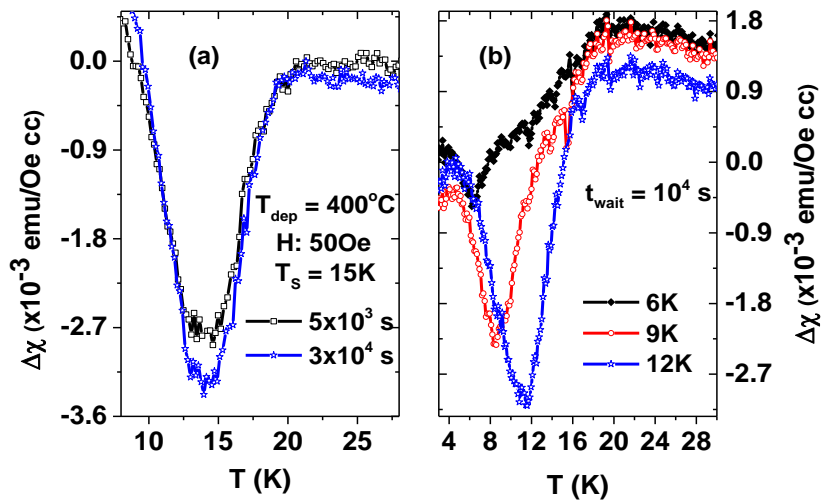

**Figure 8a:** The memory and aging effect in ZFC magnetisation shows a larger dip with increase in the wait time in a single stop and aging experiment. The system was cooled to 15 K from 300 K under zero field and it was then aged for 5 and 30 ks in separate runs before resuming the cooling to 2 K. ZFC M(T) was then measured under 50 Oe field. **Figure 8b:** The ZFC memory effect at different stop temperatures during the cooling of sample under ZFC protocol. The wait time (10 ks) was same for all the measurements. The dip in magnetisation with respect to a reference ZFC curve can be observed to occur in the close vicinity of the stop temperature.

In order to further verify the memory effect in SSG state of the sample deposited at 400 °C, magnetisation relaxation was measured as a function of time under 50 Oe field after cooling the sample to 10 K in zero field. The sample was cooled down to 10 K at 30 K/min under zero field, after which 50 Oe field was applied and magnetisation was recorded as a function of time up to $t_1$ = 5000 s. After $t_1$ seconds, the sample was quenched to 3 K under magnetic field and magnetisation was recorded for a time $t_2$ = 2000 s. At the end of time $t_2$, the sample was warmed again up to 10 K where the relaxation was measured for an additional period $t_3$ = 5000 s. The plot of magnetisation relaxation as a function of time is presented in figure 9. After application of field on ZFC sample, the magnetisation showed an instantaneous jump followed by a slow logarithmic relaxation towards a more stable configuration. At the time ($t_2$) of temporary field cooling to 3 K, the relaxation was seen to be arrested as magnetisation remained constant between $t_2$ and $t_3$. When the sample was cooled back to 10 K, the magnetisation resumed its relaxation from the same value that was attained at the end of $t_1$ and the relaxation curve during time $t_3$ was seen to be the continuation of curve during $t_1$ (inset of figure 9).

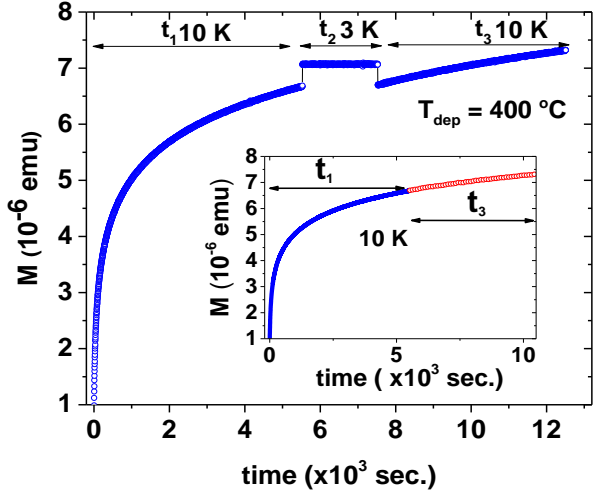

**Figure 9:** Magnetisation relaxation at 10 K in ZFC protocol with temporary cooling under field shows a logarithmic relaxation which freezes upon cooling to 3 K in 50 Oe field. The slow relaxation resumes to more stable configuration once the sample has been warmed back to 10 K. **Inset:** relaxation during time $t_3$ after warming the sample back to 10 K is the continuation of relaxation observed during time period $t_1$.

Similar aging and memory experiments were performed on the sample deposited at 300 °C and room temperature. Though 300 °C sample showed the memory effect in its SSG state, the room temperature sample could not be studied conclusively due its peak temperature value near to the minimum attainable temperature in our VSM.

## 4. Conclusion

The defect induced magnetism has been reported for the first time in reactive ion beam sputter deposited undoped AlN thin films. The films were deposited at elevated substrate temperatures. These films are a potential candidate in the dilute magnetic semiconductor research for spintronics applications. The crystallinity of the films was found to increase with increase in deposition temperature and the maximum grain size of 5-6 nm has been observed in HRTEM images. The grain density has been observed to increase with increase in deposition temperature. Nitrogen vacancy type point defects have been identified using room temperature PL measurements and are considered responsible for the spin polarisation in undoped AlN films. The detailed studies on the ZFC DC magnetisation and the frequency dependent shift in peak temperature (blocking temperature) of real component of AC susceptibility have provided ample clue to identify the possible magnetic ground states of the AlN samples. The samples deposited at T ≥ 300 °C freezes into a SSG state with critical slowing down due to strong interaction between super-spins on densely populated nano grains. Whereas, for the sample deposited at room temperature with sparsely located grains, the super-spins have been found to get blocked in a SPM state with dynamics following an Arrhenius activation behaviour. Magnetisation hysteresis has been observed in all the samples below their blocking/freezing temperatures. A thorough study of the memory and aging effect in ZFC/FC magnetisation along with time dependent relaxation in magnetisation of the sample deposited at 400 °C, unambiguously established the existence of SSG state in the nano-structured AlN thin films. The precise control over the parameters of the reactive nitrogen plasma during the reactive growth of AlN thin films can be utilized to tune the $V_N$ defect densities which in turn can be exploited for the realisation of a ferromagnetic ground state at higher temperatures in undoped AlN thin films.

**Acknowledgements**

The authors would like to thank Dr. Neha Sharma, Materials Science Group, IGCAR for providing us AlN thin film samples deposited using reactive ion beam sputter deposition system and carrying out preliminary phase characterisation. The Authors would also like to thank Dr. S. Amirthapandian for HRTEM measurements. The authors gratefully acknowledge UGC-DAE-CSR node at Kalpakkam for providing access to the 7 T cryogen free MPMS-3 SQUID VSM set up and high resolution TEM imaging.

**References**

[1] R. Frazier, G. Thaler, M. Overberg, B. Gila, C. R. Abernathy, S. J. Pearton, Indication of hysteresis in AlMnN, Applied Physics Letters 83(9) (2003) 1758-1760.

[2] D. Kumar, J. Antifakos, M. G. Blamire, Z. H. Barber, High Curie temperatures in ferromagnetic Cr-doped AlN thin films, Applied Physics Letters 84(24) (2004) 5004-5006.

[3] S. G. Yang, A. B. Pakhomov, S. T. Hung, C. Y. Wong, Room-temperature magnetism in Cr-doped AlN semiconductor films, Applied Physics Letters 81(13) (2002) 2418-2420.

[4] H. X. Liu, S. Y. Wu, R. K. Singh, L. Gu, D. J. Smith, N. Newman, N. R. Dilley, L. Montes, M. B. Simmonds, Observation of ferromagnetism above 900K in Cr–GaN and Cr–AlN, Applied Physics Letters 85(18) (2004) 4076-4078.

[5] J. Zhang, X. Z. Li, B. Xu, D. J. Sellmyer, Influence of nitrogen growth pressure on the ferromagnetic properties of Cr-doped AlN thin films, Applied Physics Letters 86(21) (2005) 212504.

[6] H. Li, H. Q. Bao, B. Song, W. J. Wang, X. L. Chen, Observation of ferromagnetic ordering in Ni-doped AlN polycrystalline powders, Solid State Communications 148(9–10) (2008) 406-409.

[7] H. Li, X. L. Chen, B. Song, H. Q. Bao, W. J. Wang, Copper-doped AlN polycrystalline powders: A class of room-temperature ferromagnetic materials, Solid State Communications 151(6) (2011) 499-502.

[8] L. -J. Shi, L. -F. Zhu, Y. -H. Zhao, B. -G. Liu, Nitrogen defects and ferromagnetism in Cr-doped dilute magnetic semiconductor AlN from first principles, Physical Review B 78(19) (2008) 195206.

[9] S. Y. Wu, H. X. Liu, L. Gu, R. K. Singh, L. Budd, M. van Schilfgaarde, M. R. McCartney, D. J. Smith, N. Newman, Synthesis, characterization, and modeling of high quality ferromagnetic Cr-doped AlN thin films, Applied Physics Letters 82(18) (2003) 3047-3049.

[10] E. Wistrela, A. Bittner, M. Schneider, M. Reissner, U. Schmid, Magnetic and microstructural properties of sputter deposited Cr-doped aluminum nitride thin films on silicon substrates, Journal of Applied Physics 121(11) (2017) 115302.

[11] A. Shah, A. Mahmood, Z. Ali, T. Ashraf, I. Ahmed, M. Mehmood, R. Rashid, I. Shakir, Influence of annealing temperature on the magnetic properties of $Cr^+$ implanted AlN thin films, Journal of Magnetism and Magnetic Materials 379 (2015) 202-207.

[12] R. Q. Wu, G. W. Peng, L. Liu, Y. P. Feng, Z. G. Huang, Q. Y. Wu, Ferromagnetism in Mg-doped AlN from ab initio study, Applied Physics Letters 89(14) (2006) 142501.

[13] J. T. Luo, Y. Z. Li, X. Y. Kang, F. Zeng, F. Pan, P. Fan, Z. Jiang, Y. Wang, Enhancement of room temperature ferromagnetism in Cu-doped AlN thin film by defect engineering, Journal of Alloys and Compounds 586 (2014) 469-474.

[14] X. H. Ji, S. P. Lau, S. F. Yu, H. Y. Yang, T. S. Herng, J. S. Chen, Ferromagnetic Cu-doped AlN nanorods, Nanotechnology 18(10) (2007) 105601.

[15] D. Pan, J. K. Jian, Y. F. Sun, R. Wu, Structure and magnetic characteristics of Si-doped AlN films, Journal of Alloys and Compounds 519 (2012) 41-46.

[16] H. H. Ren, R. Wu, J. K. Jian, C. Chen, A. Ablat, Al Vacancy Induced Room-Temperature Ferromagnetic in Un-Doped AlN, Advanced Materials Research 772 (2013) 57-61.

[17] Y. Liu, L. Jiang, G. Wang, S. Zuo, W. Wang, X. Chen, Adjustable nitrogen-vacancy induced magnetism in AlN, Applied Physics Letters 100(12) (2012) 122401.

[18] B. Song, J. C. Han, J. K. Jian, H. Li, Y. C. Wang, H. Q. Bao, W. Y. Wang, H. B. Zuo, X. H. Zhang, S. H. Meng, X. L. Chen, Experimental observation of defect-induced intrinsic ferromagnetism in III-V nitrides: The case of BN, Physical Review B 80(15) (2009) 153203.

[19] B. Fan, F. Zeng, C. Chen, Y. C. Yang, P. Y. Yang, F. Pan, Influence of strain and grain boundary variations on magnetism of Cr-doped AlN films, Journal of Applied Physics 106(7) (2009) 073907.

[20] Y. Wang, L.-T. Tseng, P. P. Murmu, N. Bao, J. Kennedy, M. Ionesc, J. Ding, K. Suzuki, S. Li, J. Yi, Defects engineering induced room temperature ferromagnetism in transition metal doped $MoS_2$, Materials & Design 121 (2017) 77-84.

[21] C.-w. Zhang, First-principles study on electronic structures and magnetic properties of AlN nanosheets and nanoribbons, Journal of Applied Physics 111(4) (2012) 043702.

[22] Neha Sharma, R. Martando, S. Ilango, T. R. Ravindran, M. S. R. Rao, S. Dash, A. K. Tyagi, Charged vacancy induced enhanced piezoelectric response of reactive assistive IBSD grown AlN thin films, Journal of Physics D: Applied Physics 50(1) (2017) 015601.

[23] Y. F. Chen, W. C. Chou, A. Twardowski, Spin-glass-like behaviour of Fe-based diluted magnetic semiconductors, Solid State Communications 96(11) (1995) 865-869.

[24] T. M. Pekarek, E. M. Watson, P. M. Shand, I. Miotkowski, A. K. Ramdas, Spin-glass ordering in the layered III-VI diluted magnetic semiconductor $Ga_{1-x}Mn_xS$, Journal of Applied Physics 107(9) (2010) 09E136.

[25] O. Mounkachi, H. El Moussaoui, R. Masrour, J. Ilali, K. El Mediouri, M. Hamedoun, E. K. Hlil, A. El Kenz, A. Benyoussef, High freezing temperature in $SnO_2$ based diluted magnetic semiconductor, Materials Letters 126 (2014) 193-196.

[26] B. Subhankar, K. Wolfgang, Supermagnetism, Journal of Physics D: Applied Physics 42(1) (2009) 013001.

[27] Neha Sharma, Shilpam Sharma, K. Prabakar, S. Amirthapandian, S. Ilango, S. Dash, A. K. Tyagi, Optical band gap and associated band-tails in nanocrystalline AlN thin films grown by reactive IBSD at different substrate temperatures, arXiv:1507.04867 (2015).

[28] C. A. Schneider, W. S. Rasband, K. W. Eliceiri, NIH Image to ImageJ: 25 years of image analysis, Nat Meth 9(7) (2012) 671-675.

[29] M. A. Signore, A. Taurino, M. Catalano, M. Kim, Z. Che, F. Quaranta, P. Siciliano, Growth assessment of (002)-oriented AlN thin films on Ti bottom electrode deposited on silicon and kapton substrates, Materials & Design 119 (2017) 151-158.

[30] Y. Sun, M. B. Salamon, K. Garnier, R. S. Averback, Memory Effects in an Interacting Magnetic Nanoparticle System, Physical Review Letters 91(16) (2003) 167206.

[31] Neha Sharma, S. Ilango, S. Dash, A. K. Tyagi, XPS studies on AlN thin films grown by ion beam sputtering in reactive assistance of $N^+/N^{2+}$ ions: Substrate temperature induced compositional variations, arXiv:1510.00541 (2015).

[32] K. Nadeem, H. Krenn, T. Traussnig, R. Würschum, D. V. Szabó, I. Letofsky-Papst, Effect of dipolar and exchange interactions on magnetic blocking of maghemite nanoparticles, Journal of Magnetism and Magnetic Materials 323(15) (2011) 1998-2004.

[33] J. L. Dormann, D. Fiorani, E. Tronc, Magnetic Relaxation in Fine-Particle Systems, Advances in Chemical Physics, John Wiley & Sons, Inc.2007, pp. 283-494.

[34] J. A. Mydosh, Spin Glasses: An Experimental Introduction, Taylor & Francis1993.

[35] P. C. Hohenberg, B. I. Halperin, Theory of dynamic critical phenomena, Reviews of Modern Physics 49(3) (1977) 435-479.

[36] M. Suzuki, S. I. Fullem, I.S. Suzuki, L. Wang, C. -J. Zhong, Observation of superspin-glass behavior in $Fe_3O_4$ nanoparticles, Physical Review B 79(2) (2009) 024418.

[37] S. Sahoo, O. Petracic, W. Kleemann, P. Nordblad, S. Cardoso, P. P. Freitas, Aging and memory in a superspin glass, Physical Review B 67(21) (2003) 214422.

[38] M. Bandyopadhyay, S. Dattagupta, Memory in nanomagnetic systems: Superparamagnetism versus spin-glass behavior, Physical Review B 74(21) (2006) 214410.

[39] M. Sasaki, P. E. Jönsson, H. Takayama, H. Mamiya, Aging and memory effects in superparamagnets and superspin glasses, Physical Review B 71(10) (2005) 104405.

[40] H. Khurshid, P. Lampen-Kelley, Ò. Iglesias, J. Alonso, M.-H. Phan, C.-J. Sun, M.-L. Saboungi, H. Srikanth, Spin-glass-like freezing of inner and outer surface layers in hollow γ-Fe$_2$O$_3$ nanoparticles, Scientific Reports 5 (2015) 15054.